# Enhanced Beam Deflection in Bent Crystals using Multiple Volume Reflection

M. B. H. Breese

*Department of Physics, National University of Singapore, Singapore 117542*

V. M. Biryukov

*Institute for High Energy Physics, Protvino, Russia*

**Abstract**

This paper presents simulations of the trajectories of high-energy ions through several bent crystal layers. At certain layer alignments volume reflection occurs from each layer and the resultant multiple volume reflection angle is correspondingly increased, along with the range of entrance angles over which ions undergo volume reflection. Another feature is that the range of entrance angles for which bent crystal channeling occurs is also increased in passing through several bent layers. The use of several bent crystal layers to produce multiple volume reflection provides an alternative approach to the design of a space shield or radiation protection at accelerators based on bent crystals.

## Introduction

Bent crystal channelling has been used for many years as a means of deflecting, extracting and collimating charged particle beams in high-energy accelerators [1-6]. If the lattice curvature radius is greater than a critical radius many ions incident on the entrance face are steered through the full curvature angle $\delta$ of the lattice and emerge aligned with the exit face. Bent crystal channelling occurs for those ions incident on the surface within the planar channeling critical angle, $\pm\theta_c$ where:

$$\theta_c = \sqrt{\frac{4Z_1Z_2e^2Nd_pCa_{TF}}{pv}} \qquad (1)$$

Here $Z_1$ and $Z_2$ are the atomic numbers of the incident and lattice nuclei, $N$ is the atomic density, $d_p$ in the planar spacing, $C \cong \sqrt{3}$, $a_{TF}$ is the Thomas-Fermi screening distance and $p,v$ are the ion momentum and velocity. The critical radius $R_c$ is given by:

$$R_c = \frac{pv}{\pi Z_1Z_2e^2Nd_p} \qquad (2)$$

To a good approximation the value of $R_c$ in silicon is given by:

$$\frac{pv}{R_c} \approx 6GeV/cm \qquad (3)$$

Volume reflection [7] occurs when the entrance beam becomes tangential to the curved lattice within the bulk of the crystal, rather than at the entrance face, resulting in reflection off the coherent field of curved lattice planes towards alignment with the entrance face. Volume reflection has been simulated for 980 GeV protons and 100 GeV/$u$ Au ions, producing a typical volume reflection angle of $\sim\theta_c$ [8]. In a recent paper [9] enhanced volume reflection produced by a single bent crystal due to the repeated passage of ions in high-energy circular accelerators was studied with the aim of reducing the local background rate for beam collimation.

Volume reflection of MeV protons along the lattice curvature of thin, bent layers was recently studied [10] using Monte Carlo simulations with the code FLUX [11]. Periodic oscillations in the volume reflected angular distribution were observed when the angular scattering for randomly aligned trajectories was less than $\theta_c$. In thicker layers no oscillations were observed, instead producing a uniform reflected angle of $\sim 0.7\theta_c$. The observations were extended to higher energies to predict when similar oscillations in the volume reflected angular distribution and nuclear encounter probability may be present in present beam extraction and collimation experiments at GeV and TeV energies.

The use of large area bent crystals as a space shield to deflect high-energy, heavy ions (HZE ions) of all atomic numbers away from spacecraft was recently proposed [12]. When the front surface of such a bent crystal shield is aligned with HZE ions, most undergo bent crystal channeling and are deflected through the full lattice curvature angle $\delta$. The bending efficiency is limited to about 75% by the planar channelling minimum yield of about 25%. This approach only works for ions which have an entrance angle at the surface of $\pm\theta_c$ to the planar bending direction.

Motivations for the present study are to provide a detailed study of the angular deflection and dispersion of high-energy ions by multiple volume reflection through several bent crystals layers and also to devise schemes capable of deflecting with high efficiency a larger range of entrance angles of HZE ions using bent crystal shields. This study is also relevant to applications on the use of bent crystals as a means of collimating and extracting beams from accelerators. Two computer codes are used to simulate 10,000 ion trajectories through several bent layers over a wide range of energies. The Monte Carlo code FLUX [11] which uses a binary collision model in conjunction with the Ziegler-Biersack-Littmark potential [13] is used to simulate ions with energies up to 100 MeV/*n* through thin layers. The code CATCH [14], based on a continuum-model [15] with Moliere potential and taking into account the single and multiple scattering on crystal electrons and nuclei, is used to simulate the passage of 400 and 980 GeV protons through much thicker layers.

Fig. 1a shows the bent crystal coordinate system used in this paper. The beam entrance angle to the channeling planes of the entrance surface is $\theta$ and the lattice is curved through an angle $\delta$. Ions which enter the surface planes within $\pm\theta_c$ undergo bent crystal channeling and most are deflected through an angle of $\delta$. Ions which enter the front surface with an angle of $-\theta$ subtended by the lattice curvature angle undergo volume reflection and are deflected by an angle up to $2\theta_c$ to a more positive exit angle. Fig. 1b shows the coordinate system used for stacking two or more curved layers together, each of which is rotated by an angle of $\Delta$ with respect to that above it. A value of $\Delta = 0.0°$ is equivalent to a single continuous layer curved through an angle of $2\delta$. For $\Delta = \delta$ the second layer has an identical angular alignment to the first layer.

## Results

The first aspect considered is the relationship between the volume reflection angle and the radius of curvature $R$ of the silicon lattice planes through which the beam passes. Fig. 2 shows the reflection angle $\theta_R$ of the transmitted beam for 5 MeV protons and 100 MeV/nucleon Au ions for different curvature radii of the (110) planes using FLUX, and for 980 GeV protons for the (111) planes using CATCH. The curvature radii are normalized to $R_c$ and $\theta_R$ normalized to $\theta_c$ to show similar trends observed over a wide range of ion energies. From Fig. 2 the ratio $\theta_R/\theta_c$ is approximately proportional to log $(R/R_c)$. For $R/R_c > 30$, almost a straight lattice, $\theta_R$ tends to constant value of 1.4-1.6$\theta_c$. $\theta_R$ decreases as $R$ decreases because the path length over which an ion travels close enough to be repelled by the lattice field is shorter. When the curvature radius becomes very small at $R/R_c \sim 1$ there no net deflection. The trend observed for heavier ions is almost identical to that for protons, so Fig. 2 demonstrates that the reflection angle for any ion type and energy can be predicted for given lattice curvature.

Fig. 3 shows the effect of a fixed lattice curvature of $R$ = 20 mm on Fe ions of different energies. Fig. 3a shows the trajectories of the randomly aligned incident ions and Fig. 3b shows the corresponding volume reflected trajectories. Randomly aligned ions undergo no net deflection whereas volume reflected ions are deflected to a more positive exit angle. The deflection is larger at lower energies, with the reflected peak shifting from +0.7$\theta_c$ to +1.5$\theta_c$ over the range of energies considered. Above an energy of 2 GeV/$n$ ($R \sim R_c$ = 16 mm) no volume reflection occurs since the lattice curvature radius is too small.

The trajectories of 100 MeV/$n$ Au ions were calculated through a silicon (110) layer ($\theta_c$ = 0.018°, $R_c$ = 0.8 mm) curved through $\delta$ = 0.08° ($R$ = 5 mm) as a function of entrance angle. Fig. 4a shows a map of the exit angular distribution on the horizontal axis for a range of entrance angles on the vertical axis. The incident beam is simulated over a wide range of entrance angles in increments of 0.002° ($\ll \theta_c$) and the exit angular distribution also sorted in steps of 0.002°. The exit angles are plotted with respect to the entrance beam direction, so undeflected ions exit at 0.0° on the horizontal axis. The general features and detailed interpretation of such maps are described in [10] and not repeated here. Bent crystal channeling occurs over a small range of entrance angles $\pm\theta_c$ , with ions bent through the lattice curvature angle. In comparison, volume reflection occurs over the full extent of entrance angles subtended by the lattice curvature, with a peak reflected angle at 0.017°, ~$\theta_c$.

Fig. 4b-d shows the effect of combining together two curved layers, each identical to that in Fig. 4a, rotated by an angle Δ. A beam portion undergoes volume reflection in both layers, resulting in a reflected peak at twice the angle produced by a single layer, i.e. ~$2\theta_c$. The exit angular distribution is broader on passing through two layers, as there are more possible combinations of interactions, with some ions reflected from both layers and others only reflected by the first or second layer. The range of entrance angles which undergo multiple volume reflection from both layers depends on the value of Δ. For Δ = δ (= 0.08°) in Fig. 4c, the exit angular distribution does not exhibit well-defined multiple volume reflection since many ions which are volume reflected by the first layer undergo bent crystal channeling in the opposite direction in the second layer. Layer structures to produce multiple volume reflection should thus be designed so that the entrance and exit surfaces of each layer are not aligned, i.e. $\Delta \neq \delta$.

In Fig. 4b,d well-defined single and multiple volume reflection occurs over a range of entrance angles subtended by the overlapping lattice curvatures. In Fig. 4b where $\Delta < \delta$ there is still a range of entrance angles where ions which are volume reflected by the first layer are deflected in the opposite direction in the second layer by bent crystal channeling. In Fig. 4d where $\Delta > \delta$ the range of entrance angles over which ions undergo multiple volume reflection is widest. The beam portion which undergoes bent crystal channelling through the first layer is deflected beyond the angular range subtended by the second layer so travels through it in a random alignment so is not deflected in the opposite sense. A layer alignment of $\Delta > \delta$ thus produces the best-defined regions with the greatest angular range of entrance angles for which multiple volume reflection occurs.

The principle of multiple volume reflection can be further extended by combining *n* curved layers together, each with $\Delta > \delta$. Fig. 5 shows the effect of combining together four and eight curved layers, each identical to those in Fig. 4. Again there are well-defined regions of entrance angles over which many ions undergo multiple volume reflection, once from each layer traversed, giving a peak which is reflected through an angle *n* times larger than that of a single layer. When the same maps in Fig. 5a,c, are plotted using an absolute exit angular scale in Fig 5b,d they are symmetrical about a diagonal lines running from the top left to the bottom right, as previously described for single curved layers [10]. This demonstrates that even such complex layer structures still obey Lindhard's reversibility rule [15].

In Fig. 6a the exit angular distributions for 100 MeV/$n$ Au ions passing through 1,2,4 identical curved layers are shown in both random and volume reflected alignment, based on Figs 4,5. The increase in the peak reflected angle with the number of layers is obvious. Fig. 6b shows exit angular distributions for eight identical curved layers, for alignments corresponding to random, multiple volume reflection and bent crystal channeling in Fig. 5. Now the peak reflected angle of $8\theta_c \sim 0.13°$ is significantly greater than the bent crystal channeling angle of $\delta = 0.08°$.

The use of several bent crystal layers also provides a way of increasing the range of entrance angles over which bent crystal channeling is produced. In Fig. 5 the range of entrance angles for which bent crystal channeling occurs is approximately $n$ times greater than that in a single layer, i.e. $2n\theta_c$. Another feature of such a multiple layer structure is that it is much less sensitive to misalignments than the use of a single curved layers, owing to the much greater range of entrance angles over which volume reflection and bent crystal channeling occur. The large angular range over which bent crystal channeling and multiple volume reflection occur results in a huge angular dispersion of the transmitted beam over a wide range of entrance angles.

Fig. 7 shows similar results to those in Fig. 6, with the trajectories of 400 GeV protons ($\theta_c = 6.3\times10^{-4}$ degrees) simulated through a layer thickness of 1 mm using CATCH. Here $\Delta > \delta$, as in Fig. 6. In Fig. 7a the peak reflected angle in passing through a single layer is $7.5\times10^{-4}$ degrees, $\sim 1.2\theta_c$, and the peak reflection angle after passing through 4 layers is $3\times10^{-3}$ degrees, i.e. four times greater. After passing through 8 layers in Fig. 7b the peak reflected angle of $6\times10^{-3}$ degrees is comparable to the bent crystal channeling angle of $\delta = 5.8\times10^{-3}$ degrees. The behaviour at high energies with CATCH is thus very similar to that observed with FLUX at lower energies, but volume reflection effects are more pronounced. This is because with increasing beam energy $E$, the random scattering angle decreases as $1/E$ while the coherent volume reflection angle decreases more slowly as $1/\sqrt{E}$, resulting in volume reflection dominating random scattering at very high energies [9]. This shows that multiple volume reflection produces highly-efficient deflection of ions with energies of many GeV to TeV per nucleon, making it an ideal mechanism for a space shield, rather than one based on bent crystal channeling as in [12].

The simulations in this paper, whether for the purpose of fabricating large-area space shield with a high deflection efficiency over a wide range of entrance angles, or for use in accelerator beam collimation rely on the availability of suitable multiple bent crystal layers. Two types of multiple layer structures can be considered. First, for lower energies a similar approach to that previously used to produce bent layers suitable for bending MeV ions can be produced by growing a graded silicon-germanium epitaxial layer [16], in which the Ge fraction varies from zero at the epilayer interface to a maximum at the surface. The resultant tetragonal strain produces a uniformly curved lattice when viewed along off-normal directions [16-18]. 10 μm layers with $R \sim 30$ mm were recently produced [19], capable of bending ions with energies up to a few GeV/*n*. This process may be extended to growing sequential curved layers, one on top of the next, so long as the cumulative lattice strain does not exceed a critical value. The second approach, more applicable to higher energies, is to align separate curved layers, whether in the form of individual bent wafers or as large area bent crystal shields as described in [12].

## Conclusions

The use of several curved layers offers the possibility of greatly expanding the range of their applications by extending both the volume reflection angle, the entrance angular arrange over which single and multiple volume reflection and bent crystal channeling occur. Similar behaviour was observed over a wide range of energies, from 5 MeV protons to 100 MeV/*n* Au to 400 GeV protons with two computer codes. Multiple volume reflection produces highly efficient deflection over a wide entrance angular range at high energies, so providing an alternative approach to the design of a space shield or radiation protection at accelerators based on curved crystals.

## Figure Captions

**Figure 1.** (a). Geometry for bent crystal channeling (BCC) and volume reflection (VR) through a lattice which is curved through an angle of $\delta$. The surface lattice planes are aligned at $\theta = 0.0°$. (b) Geometry for stacking several curved layers together to produce multiple volume reflection (MVR).

**Figure 2.** Variation of reflection angle $\theta_R$ with lattice curvature radius for different ion types and energies. The silicon layer thicknesses are 0.44 μm for 5 MeV protons, 7 μm for 100 MeV/*n* Au ions and 1 mm for 980 GeV protons.

**Figure 3.** Exit angular distributions of different energy Fe ions from a lattice curvature radius of $R = 20$ mm, layer thickness = 20 μm, i.e. $\delta \sim 0.06°$. (a) Random alignment. (b) Volume reflected alignment with the beam entrance angle half-way along the lattice curvature.

**Figure 4.** Maps showing the exit angular distributions versus entrance angle for 100 MeV/*n* Au ions from layers with $R = 5$ mm, each 7 μm thick, each curved through $\delta = 0.08°$. (a) shows a single layer, and b-d show two layers with interface rotations $\Delta$ of (b) 0.06°, (c) 0.08°, (d) 0.10°.

**Figure 5.** Maps showing the exit angular distributions versus entrance angle for 100 MeV/*n* Au ions from (a),(b) four and (c),(d) eight layers with $R = 5$ mm, each 7 μm thick, each curved though $\delta = 0.08°$ with $\Delta = 0.10°$. In (a), (c) the exit angle is plotted with respect to the entrance angle whereas in (b), (d) the exit angle is plotted in absolute terms.

**Figure 6.** Exit angular distributions with $\Delta = 0.10°$ based on Figs. 3,4. (a) Random and volume reflected entrance alignments for 1,2,4 identical curved layers. (b) Random, bent crystal channelling and volume reflected alignments for 8 identical curved layers.

**Figure 7.** Exit angular distributions for 400 GeV protons incident on silicon (111) layers. Each layer is 1 mm thick, each bent through $\delta = 5.8\times10^{-3}$ degrees with $R = 10$ m. Adjacent layers are inclined $\Delta = 6.3\times10^{-3}$ degrees to each other. (a) Random and volume reflected entrance alignments for 1,2,4 identical curved layers. (b) Random, bent crystal channelling and volume reflected alignments for 8 identical curved layers.

Figure

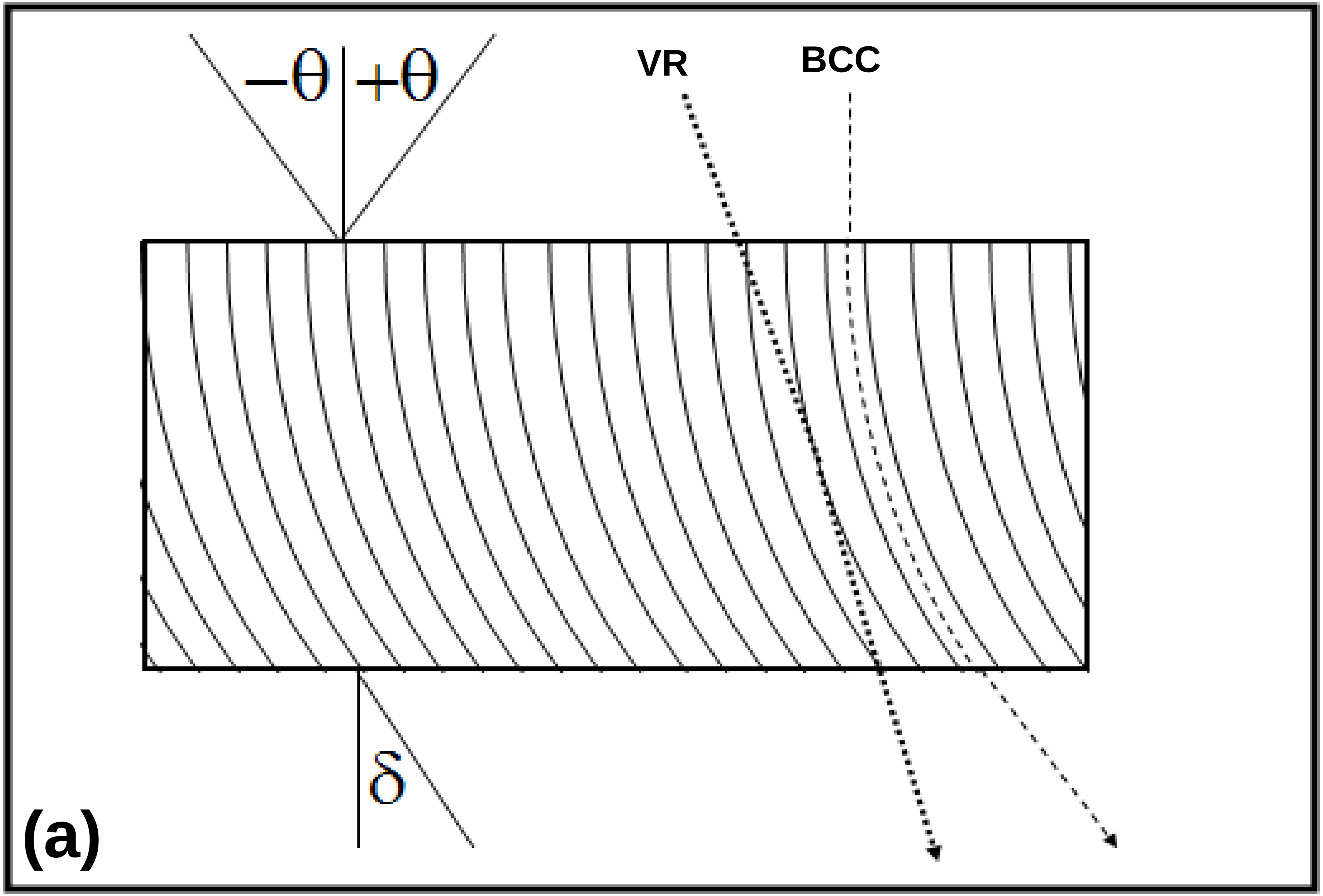

Figure 1a

Figure

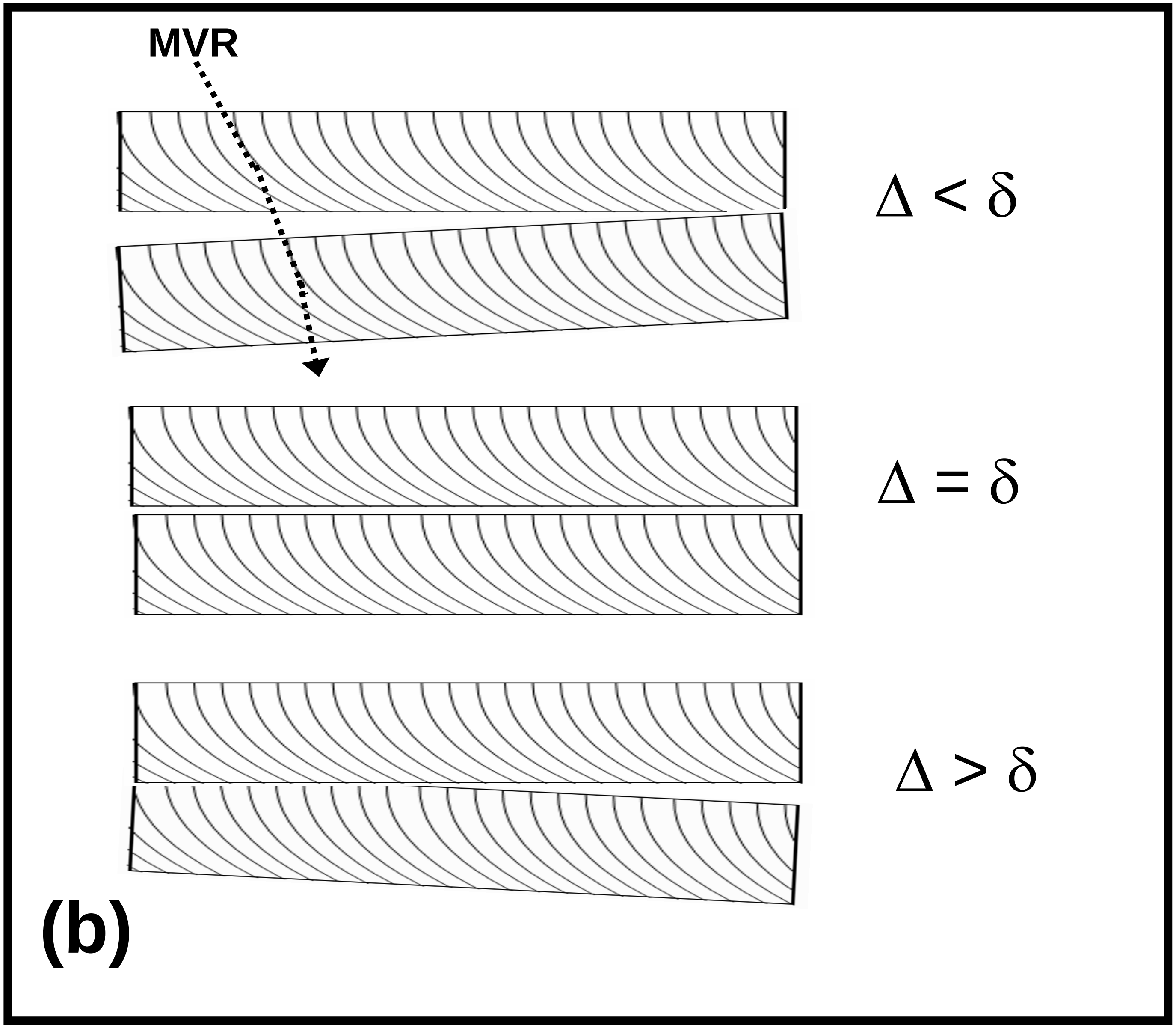

Figure 1b

Figure

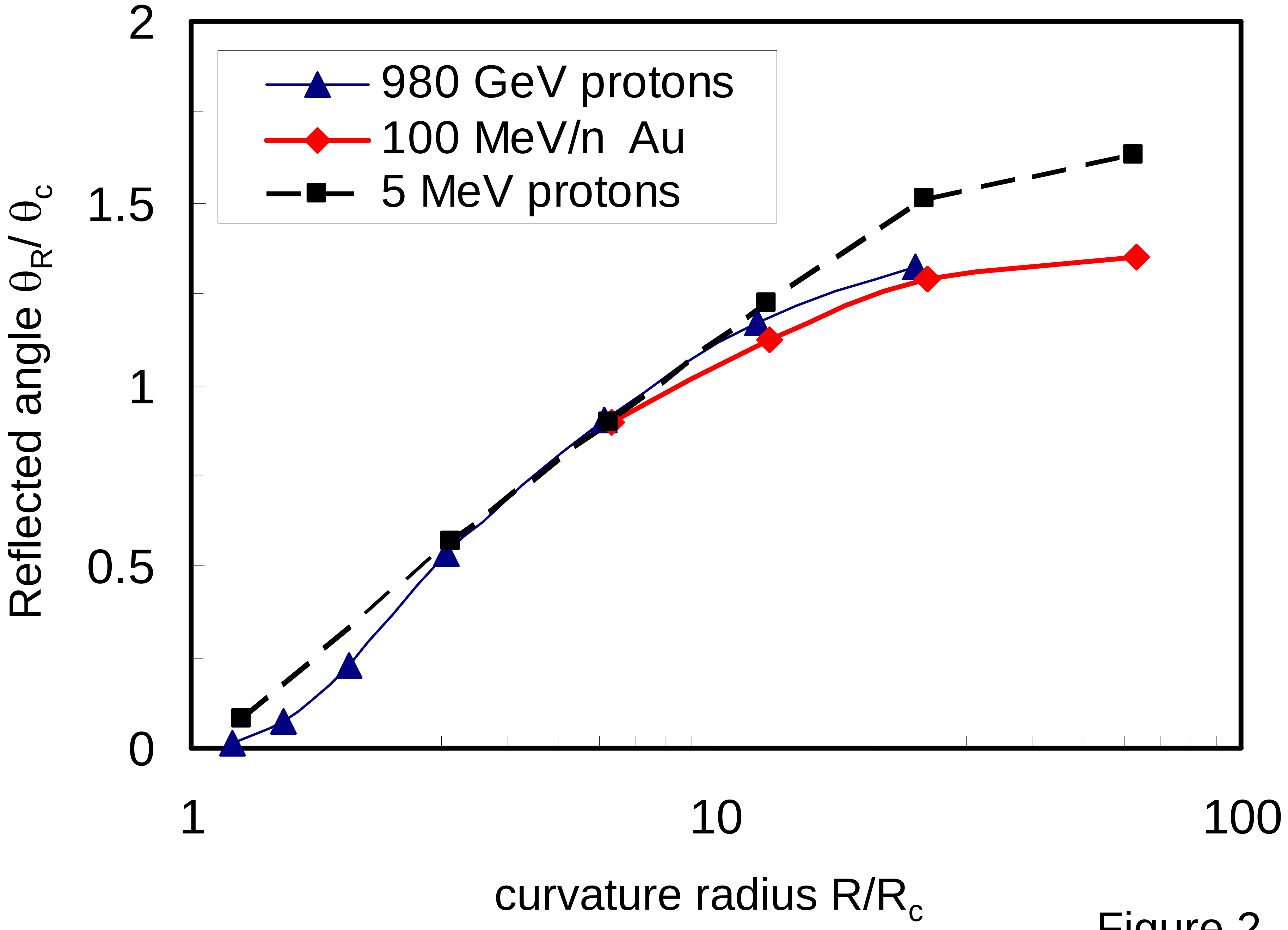

Figure 2

Figure

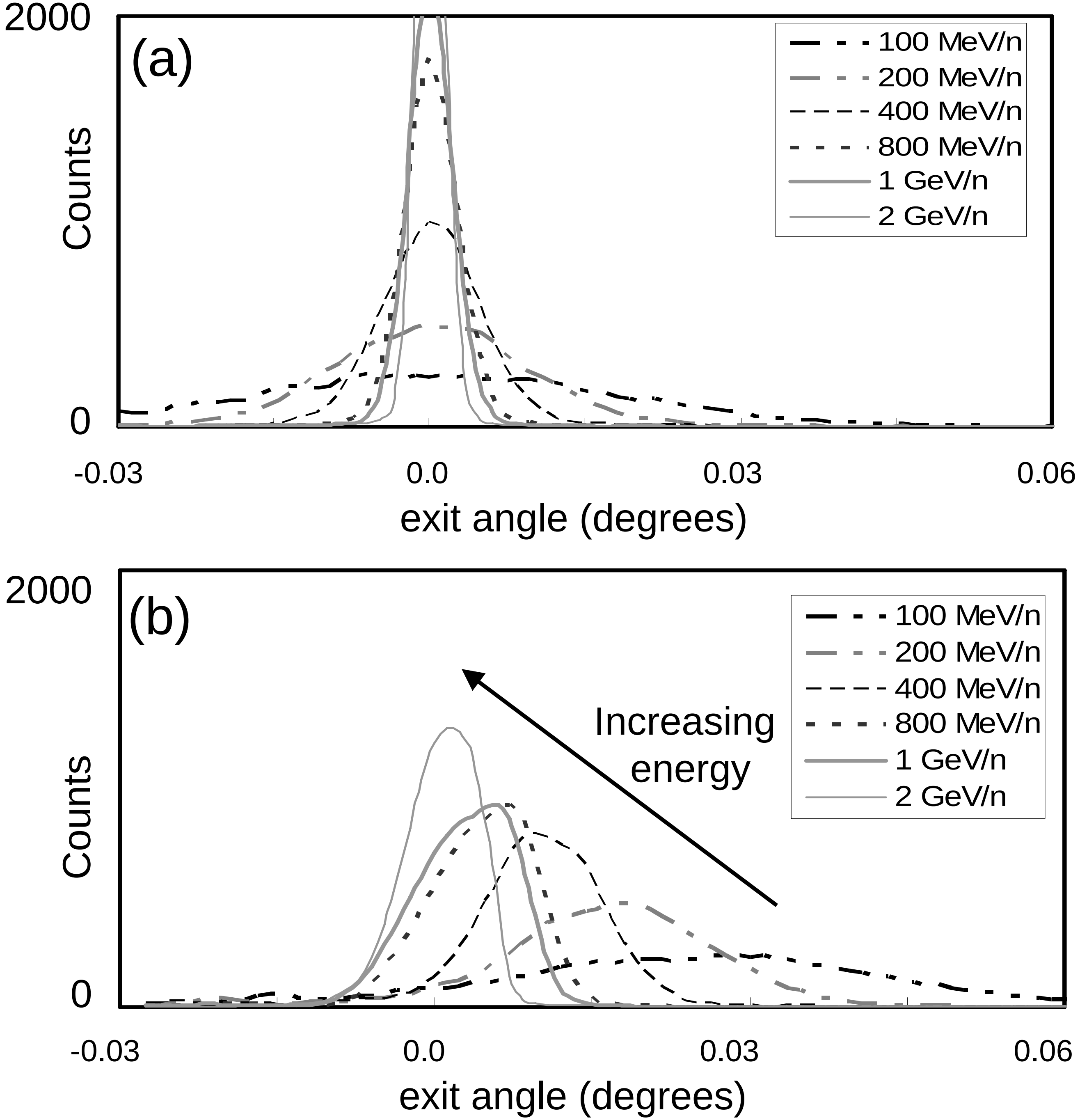

Figure 3

Figure

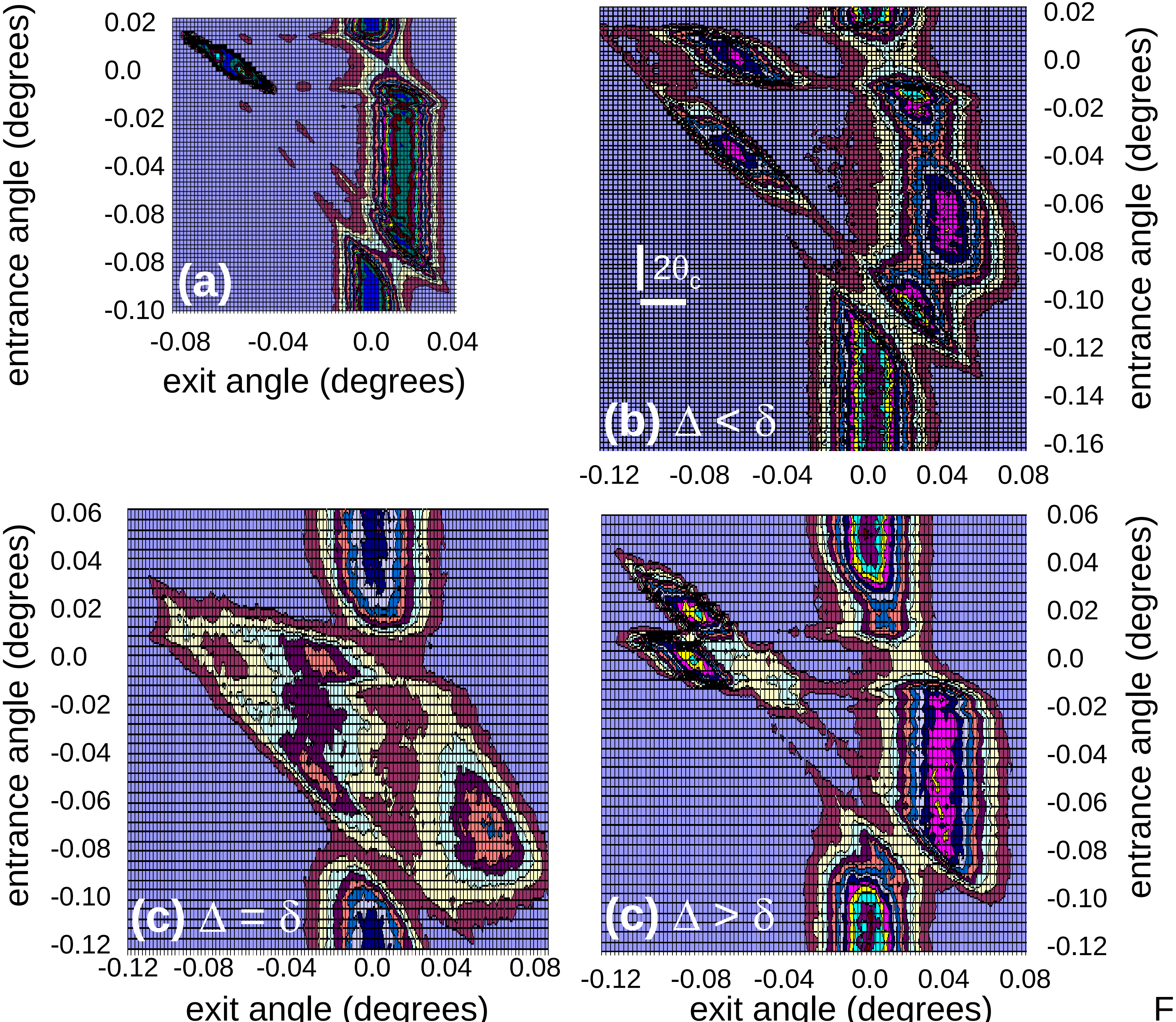

Figure 4

Figure

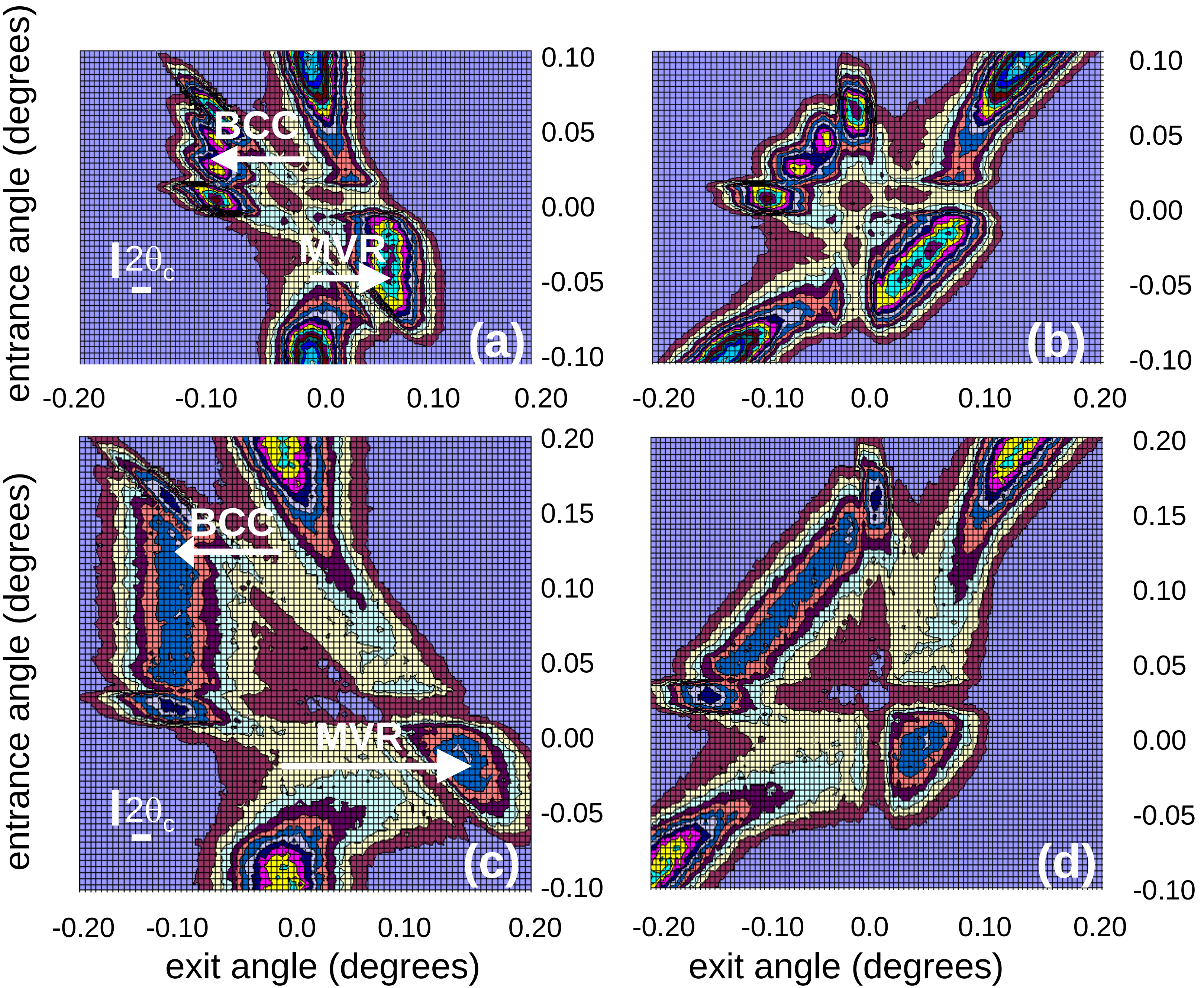

Figure 5

Figure

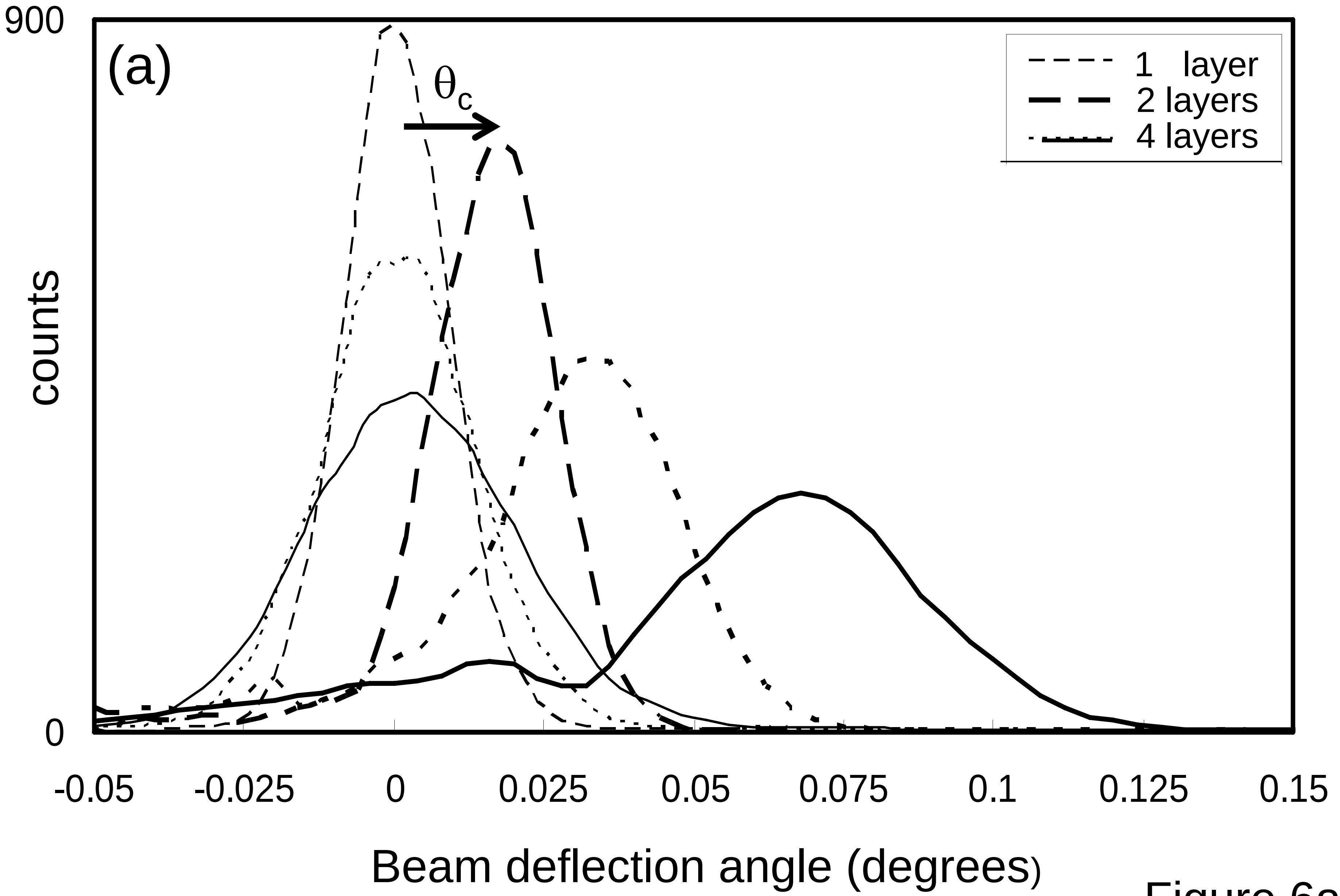

Figure 6a

Figure

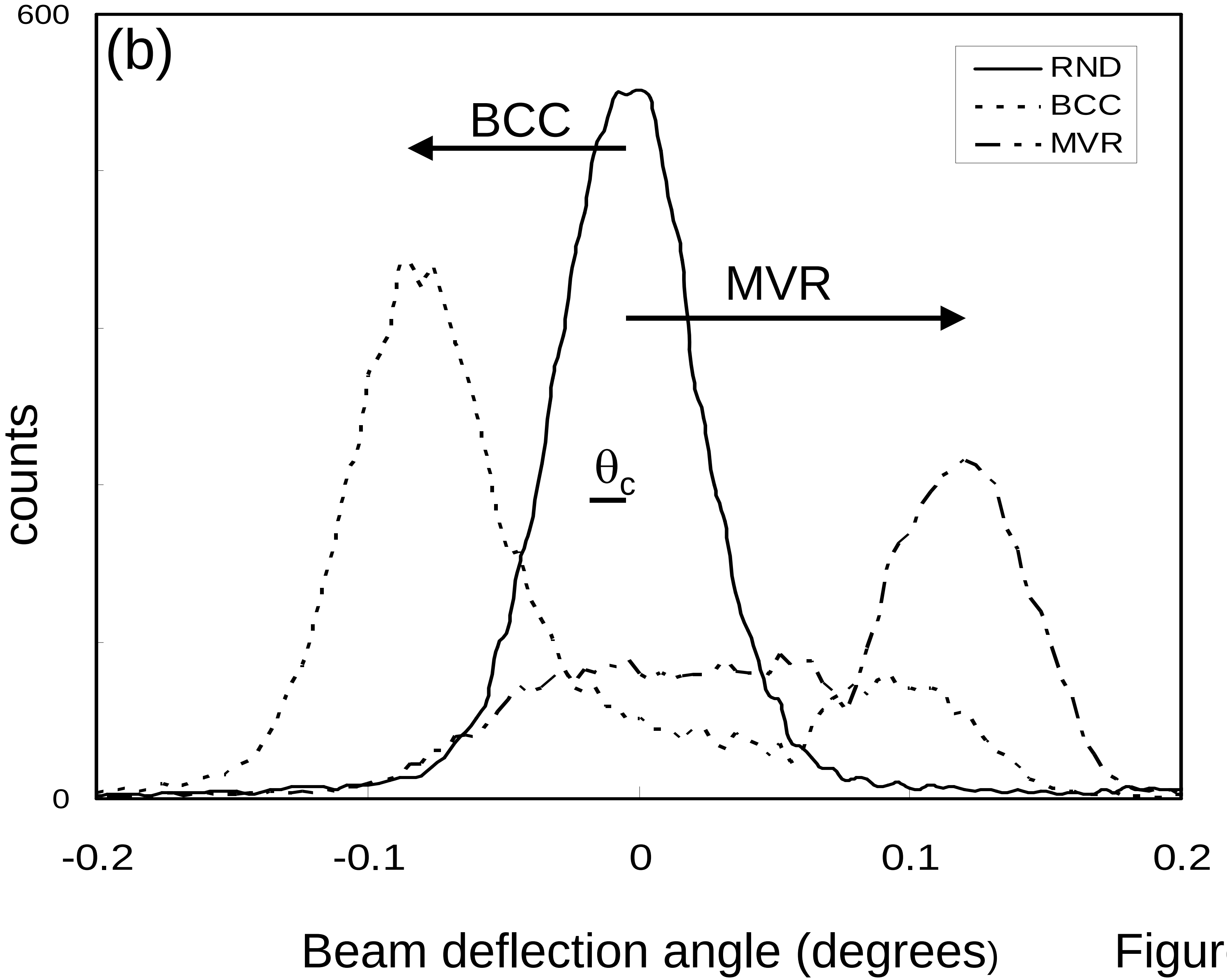

Figure 6b

Figure

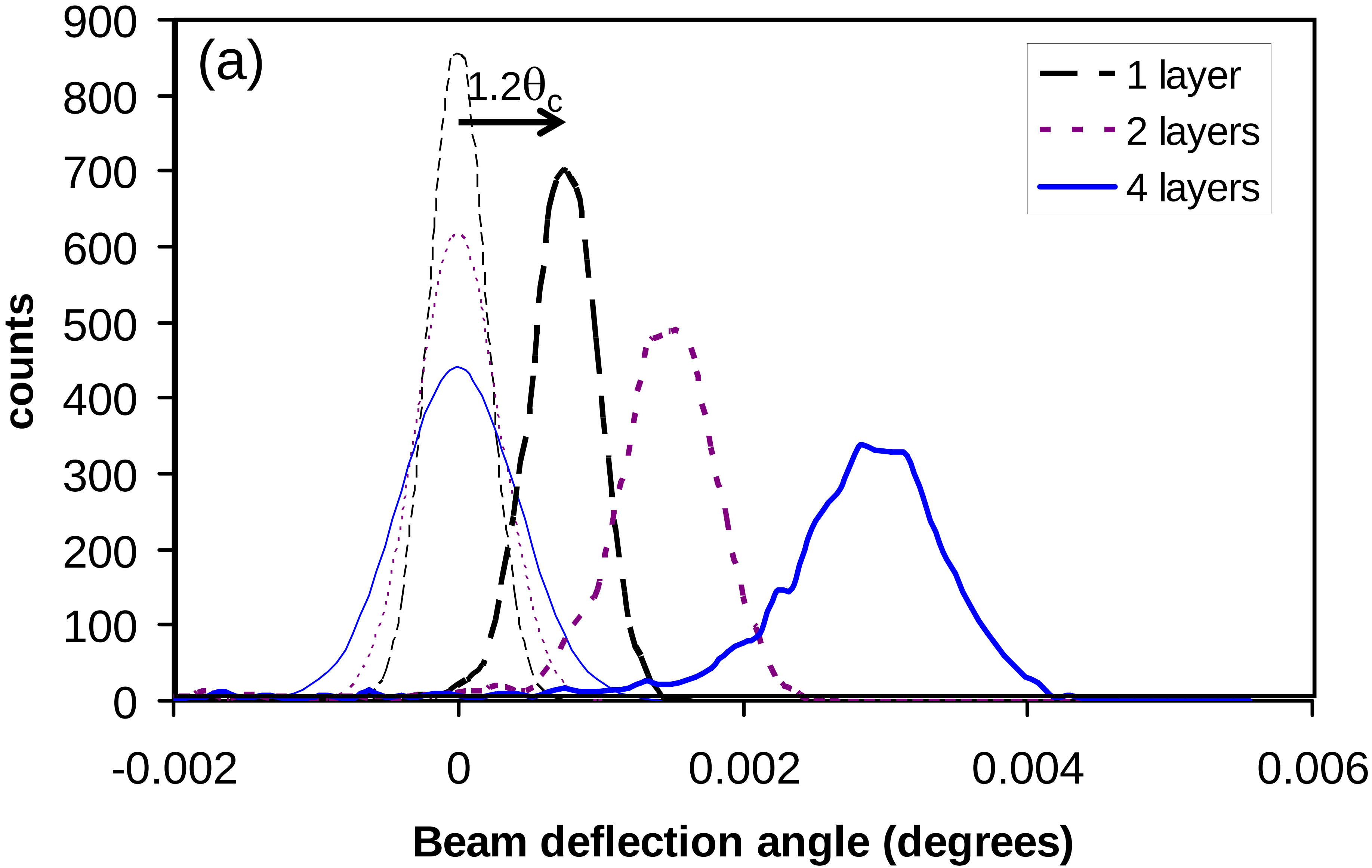

Figure 7a

Figure

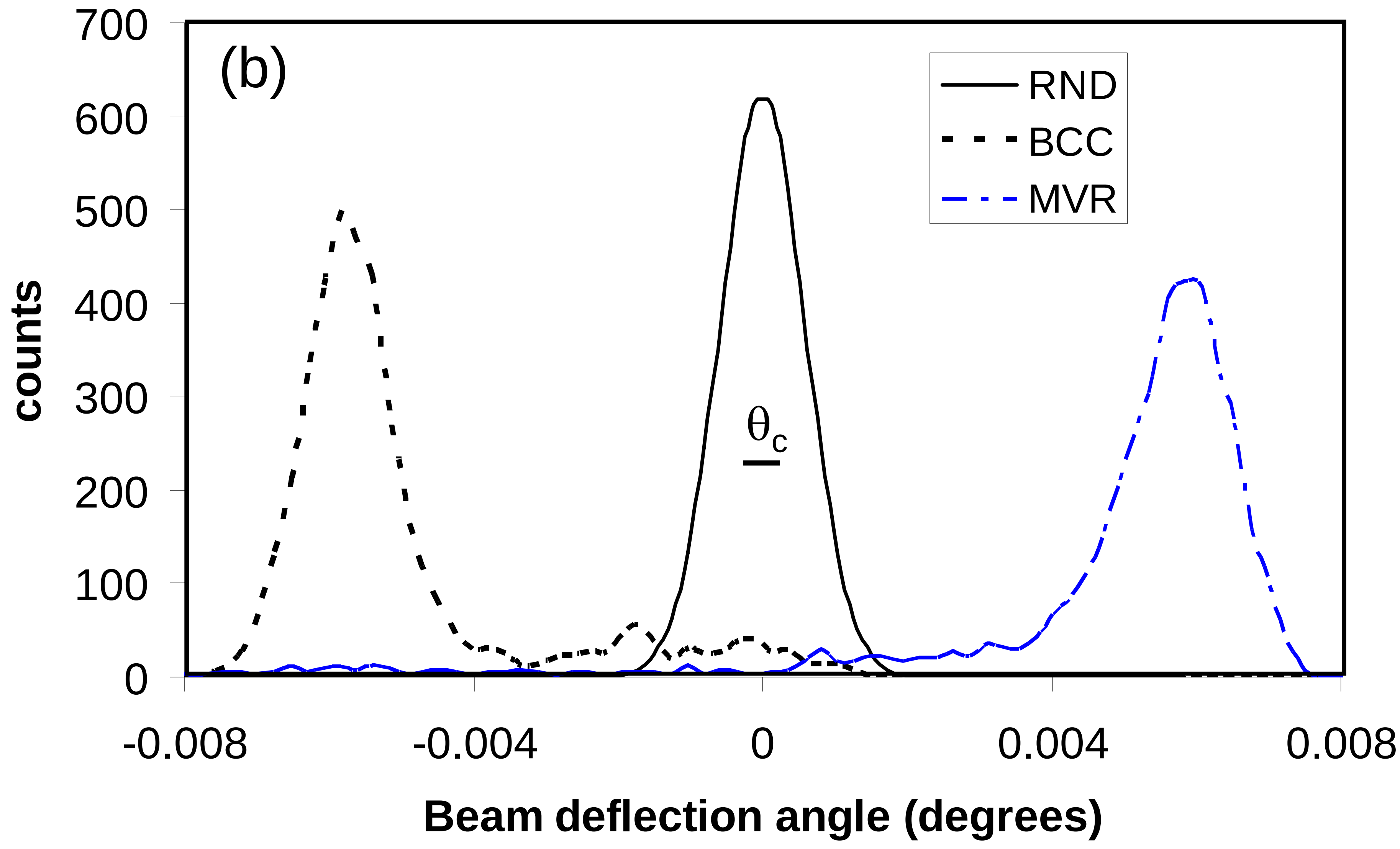

Figure 7b